\documentclass[prl,twocolumn,showpacs,showkeys,floatfix]{revtex4}

\usepackage{graphicx}
\usepackage{array}
\usepackage{amsmath,amsfonts,amssymb}

\begin{document}

\title{Kondo Effect and Surface-State Electrons}

\author{L. Limot}
\affiliation{Institut f\"{u}r Experimentelle und Angewandte Physik, Christian-Albrechts-Universit\"{a}t zu Kiel, D-24098 Kiel, Germany}
\author{R. Berndt}
\affiliation{Institut f\"{u}r Experimentelle und Angewandte Physik, Christian-Albrechts-Universit\"{a}t zu Kiel, D-24098 Kiel, Germany}
%\date{\today}

\begin{abstract}
We have used low temperature scanning tunneling spectroscopy and atomic manipulation to study the role of
surface-state electrons in the Kondo effect of an isolated cobalt atom adsorbed on Ag(111).
We show that the observed Kondo signature remains unchanged in close proximity of a monoatomic step, where
the local density of states of the surface-state electrons is strongly perturbed. This result indicates a minor
role for surface-state electrons in the Kondo effect of cobalt, compared to bulk electrons. A possible
explanation for our findings is presented.
\end{abstract}



\maketitle
Recently, scanning tunneling microscopy (STM) has been used to investigate the electronic properties
of single magnetic impurities on the surface of metals, opening a new avenue of research into the Kondo effect \cite{Li,Madhavan}.
The Kondo effect occurs when a magnetic impurity in a metal couples to surrounding conduction electrons.
This coupling, which involves spin-flip scattering events on the site of the impurity, causes conduction electrons
to form a correlated ground state that screens the magnetic moment of the impurity at temperatures below a
characteristic Kondo temperature ($T_{K}$) \cite{Kondo HF}. It also leads to a narrow resonance in the
local density of states (LDOS) of the  impurity at the Fermi energy ($E_{F}$), which is detected in STM and, in particular,
in scanning tunneling spectroscopy (STS).

Owing to the atomic resolution in space and the better than meV resolution in energy, STM and STS
can perform local investigations of the Kondo effect, directly at the adsorption site
of single magnetic atoms on the surface of metals. However, for a comparison to bulk measurements,
differences in the bulk and in the surface electronic structures must be taken into account. For instance,
well known bulk Kondo systems yield Kondo temperatures
substantially higher than the ones reported by STM \cite{Madhavan,Manoharan,Knorr}. Furthermore, most STM and STS
measurements have been performed on (111) facets of noble metals, where bulk electrons co-exist with Shockley surface-state electrons, which form a
quasi-two-dimensional electron gas trapped between the surface barrier potential and a band gap in the
crystal \cite{SurfState}.

So far, there still is a lack of experimental and theoretical data concerning the role of the surface-state
electrons in the Kondo effect. In this brief report we specifically focus on this topic. We report a STM and
a STS study for cobalt adatoms on Ag(111), where the combination of controlled cobalt manipulation over the surface
and tunneling-spectroscopy of cobalt adatoms enables us to tune the LDOS of the surface-state electrons and
thus probe its influence on the Kondo temperature of Co/Ag(111). We also provide a possible scenario to interpret
our data, which, we hope, will trigger further theoretical studies.

\begin{figure}[t]
\includegraphics[width=4.5cm,bbllx=100,bblly=520,bburx=345,bbury=785,clip=]{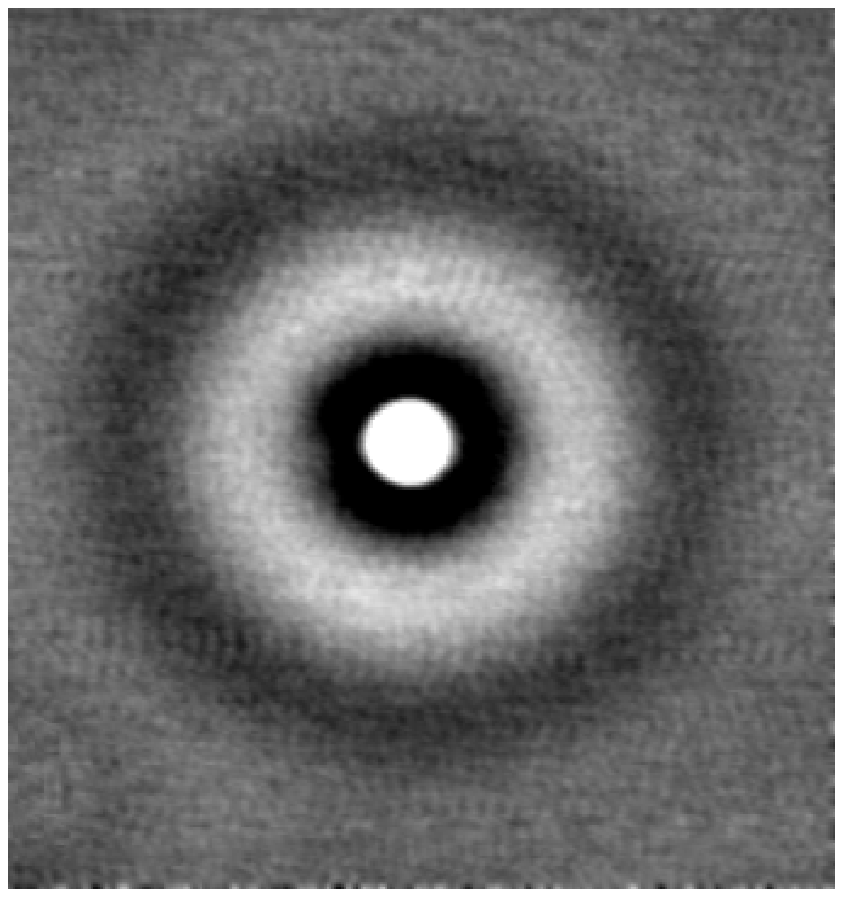}
\caption{Constant current STM image of an isolated cobalt adatom on Ag(111) ($11\times 11$ nm$^{2}$,
$I=0.5$ nA, $V=100$ mV). Also visible, the scattering of the Ag(111) surface state by the adatom.}
\label{fig1}
\end{figure}

The measurements were performed in a home-built ultrahigh vacuum STM at a working temperature of $T=4.6$ K and
the Ag(111) surface was cleaned by Ar$^{+}$ sputter/anneal cycles. Single cobalt atoms were evaporated onto the
cold Ag substrate by heating a degassed cobalt wire ($>99.99\%$) wound around a pure W wire ($>99.95\%$). The evaporation,
through an opening of the He shield of the cryostat, yielded a coverage of $3\times 10^{-3}$ ML. No appreciable
increase of other impurities was detected. After the evaporation, the thermalization of Ag(111) to $4.6$ K
ensured a negligible thermal diffusion of cobalt adatoms within measurement times. Spectra of the differential
conductance of the tunneling junction, $dI/dV(V)$, were recorded via lock-in detection (AC modulation was $1$ mV (rms) in amplitude
and $\sim 10$ kHz in frequency), where $V$ is the sample bias measured with respect to the tip. The images and the
spectra were recorded with a etched tungsten tip; the tip was further treated \textit{in situ} by soft indentations into the
Ag(111) surface, until the cobalt adatoms were imaged spherically (Fig.~\ref{fig1}), and the $dI/dV$ had no structure
near the zero bias voltage, i.e. the Fermi energy. The cobalt adatoms have then a Gaussian-like profile $\sim 0.7$
$\text{\AA}$ high with an apparent diameter of $\sim 6$ $\text{\AA}$ (full width at half maximum).

\begin{figure}[b]
\includegraphics[width=8.0cm,bbllx=30,bblly=520,bburx=580,bbury=780,clip=]{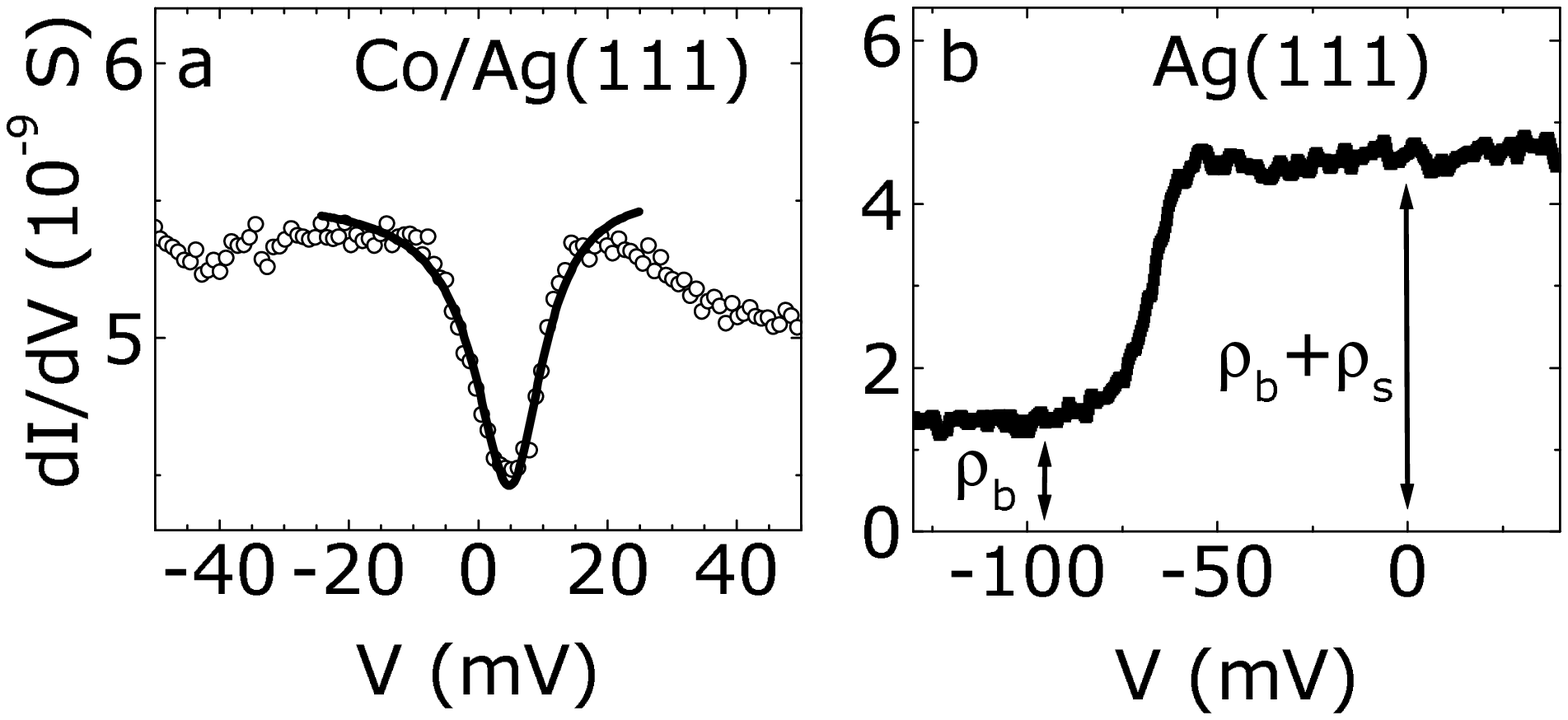}
\caption{a) $dI/dV$ spectrum taken at the center of a cobalt adatom ($I=0.5$ nA, $V=100$ mV). The spectrum is
an average of $10$ single spectra from varying cobalt adatoms and tips. Solid line: Fano fit
following the expression of Ref.~\onlinecite{Ujsaghy}. b) $dI/dV$ spectrum taken over a bare terrace of Ag(111)
($I=0.5$ nA, $V=100$ mV). The onset at $-67$ meV corresponds to the energy of the low band edge of the Ag(111)
surface-state. After the onset, the contribution of the surface LDOS to the $dI/dV$ with respect to
the bulk LDOS is $\rho_{s}/\rho_{b}\approx 2$.}
\label{fig2}
\end{figure}

A typical $dI/dV$ spectrum acquired at the center of a cobalt adatom is presented in Fig.~\ref{fig2}a.
The dip in the LDOS near $E_{F}$ is the signature in STS of the Kondo effect of Co/Ag(111).
When the tip is moved laterally off the cobalt adatom, the dip continuously decreases in amplitude and vanishes
at $\approx 8$ $\text{\AA}$ away from the center of the cobalt atom; at a distance of $\agt 50$ $\text{\AA}$,
the $dI/dV$ spectrum of the Ag(111) surface-state is recovered (Fig.~\ref{fig2}b). A similar dip at $E_{F}$
has previously been observed in STS for other Kondo systems like Ce/Ag(111) and Co/Cu(111) \cite{Li,Manoharan,Knorr},
in contrast to the asymmetric line shape observed for Co/Au(111), Ti/Ag(100) and Co/Cu(100) \cite{Madhavan,Nagaoka,Knorr}.
The $dI/dV$ spectrum can be understood following the pioneering framework by Fano \cite{Fano}. When $T\ll T_{K}$,
spin-flip processes are frozen out and the Kondo many-body problem reduces to a single-particle
resonance near $E_{F}$ of width $2k_{B}T_{K}$ \cite{Noziers,Kondo HF}, which, in principle,
should be detected in a $dI/dV$ spectrum. However, tunneling from the tip into the Kondo resonance occurs
through two channels: a direct one, and an indirect one via
the $sp$ substrate conduction band locally hybridized with the cobalt $d$ levels (which are
giving rise to the magnetic moment of the cobalt atom). The interference between the two channels $-$
known as Fano interference $-$ produces a symmetric or an asymmetric line in the $dI/dV$ spectrum,
depending on the relative weight of the two channels. Modelling of the dip of Fig.~\ref{fig2}a with
a Fano line \cite{Ujsaghy,Ellipse2}, allows then to extract from the $dI/dV$ spectrum information concerning the Kondo effect
for Co/Ag(111). The Fano fits on various cobalt spectra yielded the following results:
from the line width we extract a temperature $T_{K}=83(10)$ K higher than our working temperature of 4.6 K,
a Fano parameter $q=0.0(1)$ $-$ indicating that the predominant tunneling channel is through the $sp$ band $-$
and a shift of the resonance to $5.8(4)$ meV above $E_{F}$. The reported values are in good agreement with a
previous STS study of Co/Ag(111), where a temperature of $T_{K}=92(6)$ K was found by Schneider \textit{et al.}
\cite{Schneider}.

\begin{figure}[b]
\includegraphics[width=8.0cm,bbllx=70,bblly=515,bburx=570,bbury=780,clip=]{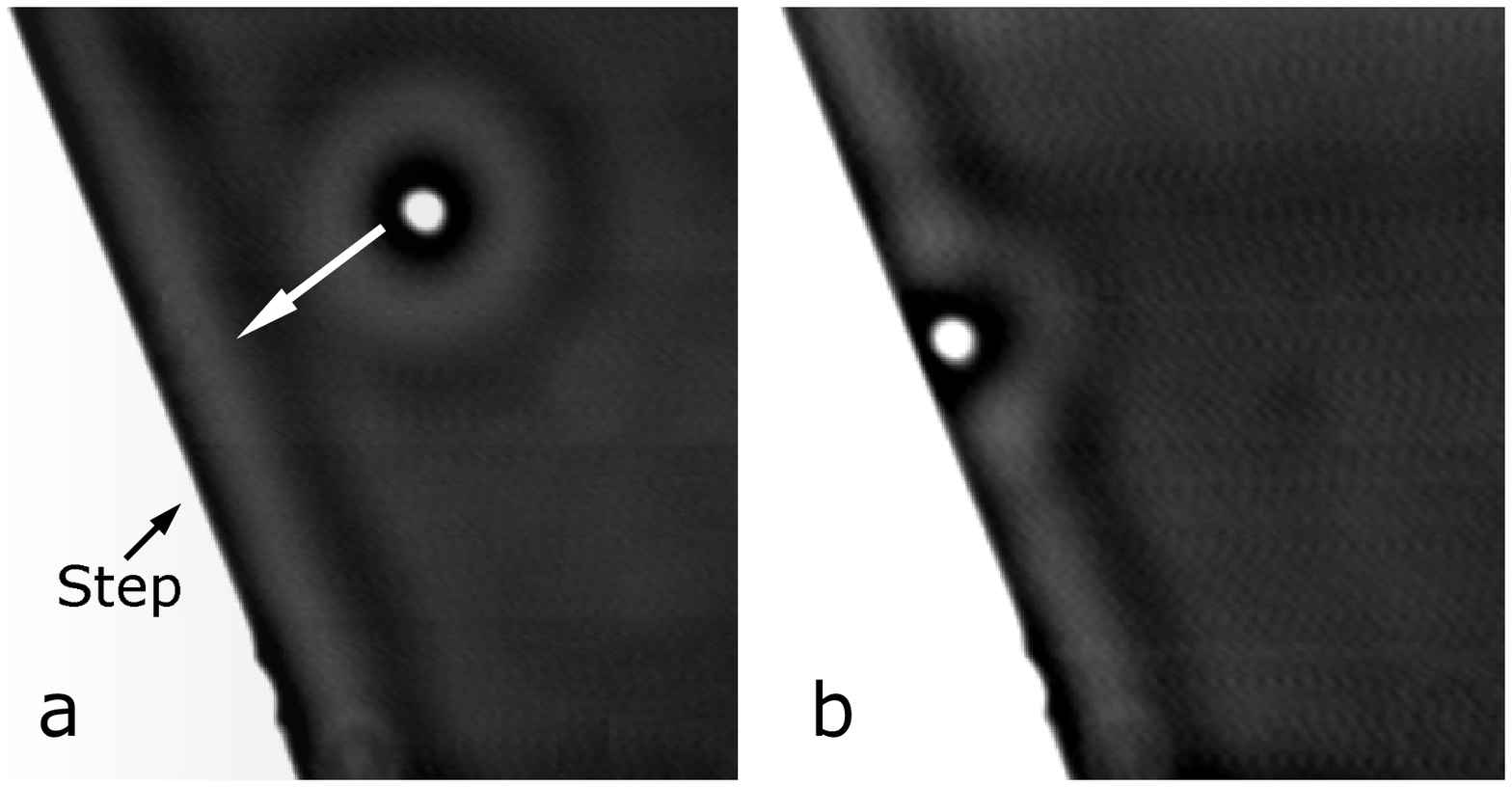}
\caption{Cobalt adatom near a monoatomic step of Ag(111) ($17\times 17$ nm$^{2}$, $I=0.5$ nA, $V=10$ mV). a) Prior, and,
b) After manipulation.}
\label{fig3}
\end{figure}

\begin{figure}[t]
\includegraphics[width=5.0cm,bbllx=30,bblly=275,bburx=320,bbury=800,clip=]{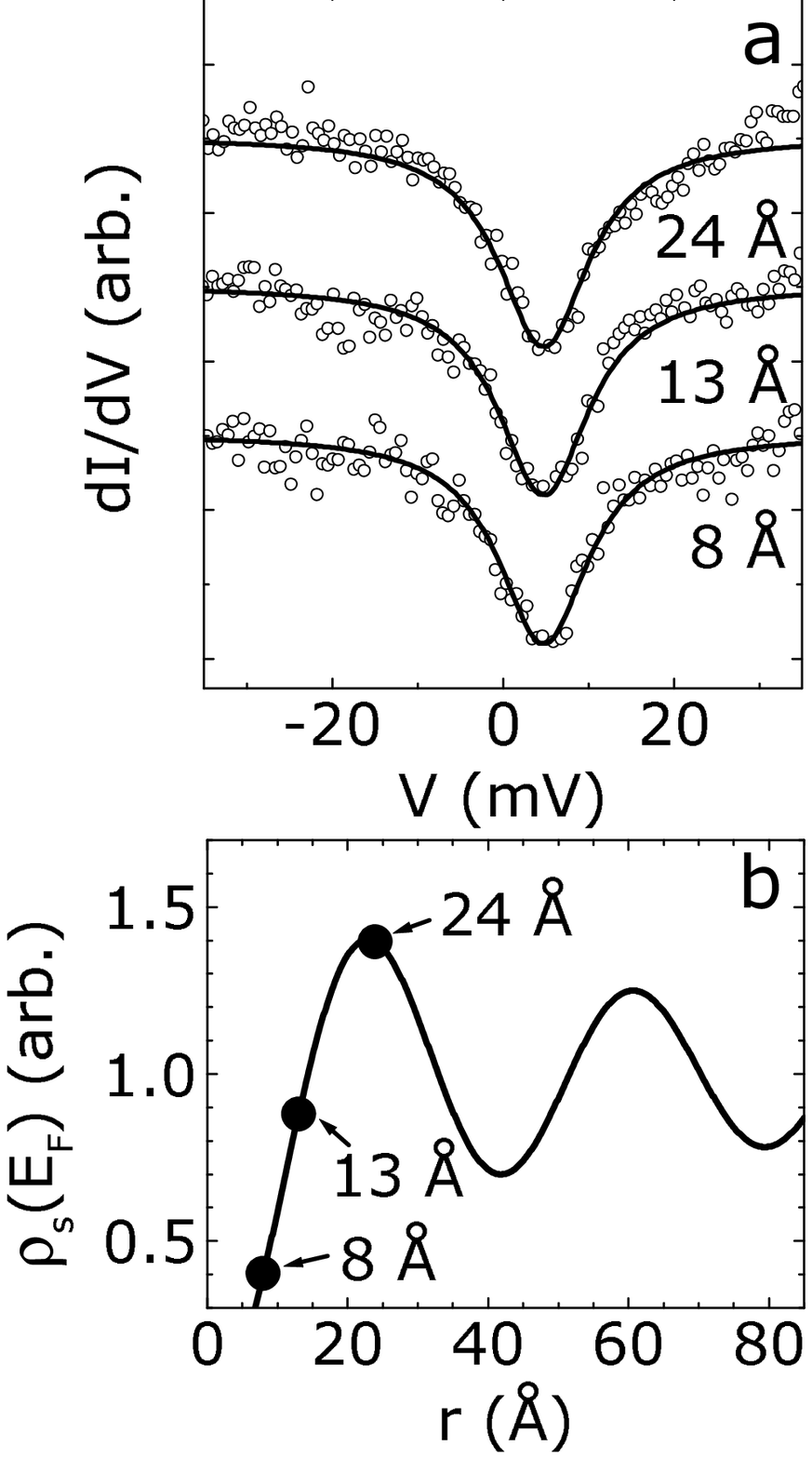}
\caption{a) $dI/dV$ spectra of a cobalt adatom at $24$, $13$ and $8$ $\text{\AA}$ from the bottom
edge of a monoatomic step of Ag(111), and Fano fits to the spectra (solid lines). The spectra, shifted vertically
for clarity, were acquired by opening the feedback loop at $I=0.5$ nA and $V=100$ mV. b) Calculated
spatial dependence of $\rho_{s}(E_{F})$ near a step, following Refs.~\onlinecite{SurfState}
and \onlinecite{LDOS step} ($r$ is the distance from the step edge). The solid circles indicate the positions were
the spectra of Fig.~\ref{fig4}a were acquired.}
\label{fig4}
\end{figure}

Figures~\ref{fig3} and \ref{fig4} illustrate the main experimental findings of this report. To explore the role
played by the Ag(111) surface-state electrons in the Kondo effect of Co/Ag(111), we combined manipulation and
spectroscopy experiments to probe how the Kondo effect of Co/Ag(111) is affected in close proximity
of a monoatomic step of Ag(111) where the surface-state LDOS ($\rho_{s}$) is strongly perturbed.
Figure~\ref{fig3} illustrates a typical manipulation procedure we performed. By placing the STM tip over a
cobalt adatom and by applying $I=2.8$ nA and $V=2.8$ V to the tunneling junction (Fig.~\ref{fig3}a),
the cobalt adatom is trapped below the tip and gently dragged to a desired location on the surface
$-$ in Fig.~\ref{fig3}b, at $8$ $\text{\AA}$ from a
monoatomic step of Ag(111). Figure~\ref{fig4}a presents $dI/dV$ data acquired over a cobalt adatom positioned
below the step of Fig.~\ref{fig3}, at distances $r$ from the step edge of $8$, $13$ and $24$ $\text{\AA}$.
The Fano fits to the three spectra do not reveal any appreciable change in the resonance, indicating that $T_{K}$ is
constant with $r$. Near a step, however, $\rho_{s}$ is strongly perturbed at $E_{F}$ because of the back-scattering
of the surface state by the step $-$ contrary to the bulk
LDOS ($\rho_{b}$). Neglecting thermal broadening, the spatial variation of $\rho_{s}(E_{F})$ is
approximately described by a Bessel function $1-J_{0}(2k_{F}r)$ \cite{SurfState,LDOS step},
where $k_{F}=0.084$ $\text{\AA}^{-1}$ for Ag(111). The dependency of $\rho_{s}(E_{F})$ on $r$ is
presented in Fig.~\ref{fig4}b where we have also indicated the positions where the spectra of Fig.~\ref{fig4}a
were recorded. As shown, when moving from $r=24$ $\text{\AA}$ to $r=8$ $\text{\AA}$, $\rho_{s}(E_{F})$ decreases
by more then a factor three, to be compared to the constant value of $T_{K}$ observed for Co/Ag(111) at a Ag(111)
step. The only significant spectroscopic change we observe in the Kondo effect of Co/Ag(111)
is for artificially fabricated cobalt dimers, where, in agreement with the
findings of Ref.~\onlinecite{Chen}, an abrupt disappearance of the Kondo resonance occurs.

To summarize the experimental data, we have studied the Kondo effect of cobalt atoms adsorbed on the Ag(111) surface,
which also exhibits surface-state electrons. The STS data acquired near a Ag(111) step indicate that the Kondo
temperature of Co/Ag(111) is robust on substantial changes in the LDOS of the surface-state electrons,
suggesting a minor role for these electrons in this Kondo system.

To discuss these findings, we recall some fundamental aspects of the Kondo problem.
The Kondo effect is essentially determined by one energy scale: $k_{B}T_{K}$. Within the Coqblin-Schrieffer
model \cite{Kondo HF}, the Kondo temperature is
\begin{equation}
T_{K}=D e^{-1/2J\rho},
\label{Kondo}
\end{equation}
where $J$ is the antiferromagnetic coupling strength between conduction electrons and the magnetic moment of a cobalt
atom, and $\rho$ the LDOS at $E_{F}$ ($D$ is the bandwidth of silver). Although a quantitative interpretation
of our experimental findings would need a microscopic theory describing the Kondo scattering of bulk
and of surface-state electrons, which is beyond the scope of this report, we may draw, based
on Eq.~(\ref{Kondo}), some conclusions concerning the involvement of surface-state electrons in the Kondo effect.

First, as seen in Eq.~(\ref{Kondo}), the Kondo temperature of a magnetic scatterer depends
exponentially on $J\rho$. Since experimentally $T_{K}$ is found to be insensitive to a variation
of $\rho_{s}$ by a factor of three, we conclude that the Kondo effect of cobalt atoms adsorbed on Ag(111) involves
mainly bulk electrons: $J\rho\approx J_{b}\rho_{b}$. This conclusion agrees with those of
Ref.~\onlinecite{Knorr}, where the Kondo temperature of the Co/Cu(111) and Co/Cu(100) systems was shown
to vary with the number of next nearest Cu neighbors ($n$) of the cobalt atom, hence with $\rho_{b}\sim n$.

Next, we focus more specifically on the surface-state electron coupling $J_{s}$.
At the tip-surface distances where usual $dI/dV$ spectra are acquired ($\sim 10$ \text{\AA} over the surface),
about 2/3 of the current into the Ag(111) surface is due to tunneling into the surface-state LDOS (see Fig.~\ref{fig2}b),
and in particular $\rho_{s}/\rho_{b}\approx 2$ at $E_{F}$. Assuming that this ratio between the two LDOS holds also at
the cobalt site, i.e. that $\rho_{s}\sim\rho_{b}$ at the impurity site, the feeble contribution of the surface-state electrons to the Kondo effect is to be found then in a
weaker coupling to the magnetic moment of the cobalt atom compared to bulk electrons: $J_{s}\ll J_{b}$.
While $J_{s}$ is small, it is clearly non-zero, as demonstrated
by Manoharan \textit{et al.} in their quantum mirage experiment \cite{Manoharan}, where they
exploited the scattering of the Cu(111) surface-state electrons in an elliptical quantum corral
to project the Fano line of a cobalt adatom in one focus to the other empty focus, about $80$ $\text{\AA}$ away.
However, only ellipses of specific dimensions will have a sufficient surface-state electron
amplitude at the focal adatom to yield a detectable Kondo signal at the opposite focus \cite{Ellipse,Ellipse2}, in
agreement with our finding of a minor involvement of surface-state electrons in the surface Kondo effect.

In conclusion, we have studied the Kondo effect of cobalt atoms adsorbed on the surface of Ag(111) by STM
and by STS. The Kondo signature in the $dI/dV$ of Co/Ag(111) is a dipped Fano resonance near $E_{F}$ from which
is extracted a Kondo temperature of $T_{K}=83(10)$ K. A combination of cobalt-manipulation and tunneling spectroscopy,
suggests, on an experimental basis, that the role of surface-state electrons in the Kondo effect of Co/Ag(111)
is minor, possibly because of a weaker coupling $J_{s}$ compared to the one of the bulk electrons.
This appears to be a general property of Kondo systems and, we hope, will trigger further theoretical
studies for a quantitative understanding of our experimental data.

We gratefully acknowledge A. Schiller and R. Bulla for fruitful discussions, and
thank the Deutsche Forschungsgemeinschaft for financial support.


\begin{thebibliography}{99}

\bibitem{Li} J. Li, W.-D. Schneider, R. Berndt, and B. Delley, \prl 80  (1998) 2893.

\bibitem{Madhavan} V. Madhavan, W. Chen, T. Jamneala, M. F. Crommie, and N. S. Wingreen, Science
280 (1998) 567.

\bibitem{Kondo HF} A. C. Hewson, The Kondo Problem to Heavy Fermions (Cambridge University Press,
Cambridge, England, 1993).

\bibitem{Manoharan} H. C. Manoharan, C. P. Lutz, and D. M. Eigler, Nature (London) 403, (2000) 512.

\bibitem{Nagaoka} K. Nagaoka, T. Jamneala, M. Grobis, and M. F. Crommie, \prl 88, (2002) 077205.

\bibitem{Knorr} N. Knorr, M. A. Schneider, L. Diekh{\"o}ner, P. Wahl, and K. Kern, \prl 88, (2002) 096804.

\bibitem{SurfState} L. C. Davis, M. P. Everson, R. C. Jaklevic, and W. Shen, \prb 43 (1991) 3821;
Y. Hasegawa and P. Avouris, \prl 71 (1993) 1071.

\bibitem{Fano} U. Fano, Phys. Rev. 124 (1961) 1866.

\bibitem{Noziers} P. Nozieres, J. Low Temp. Phys. 17 (1974) 31.

\bibitem{Ujsaghy} O. {\'U}js{\'a}ghy, J. Kroha, L. Szunyogh, and A. Zawadowski,
\prl 85 (2000) 2557.

\bibitem{Schneider} M. A. Schneider, L. Vitali, N. Knorr, and K. Kern, \prb 65 (2002) R121406.

\bibitem{Chen} W. Chen, T. Jamneala, V. Madhavan, and M. F. Crommie, \prb 60 (1999) R8529.

\bibitem{LDOS step} J. Li, W.-D. Schneider, and R. Berndt, \prb 56 (1997) 7656;
O. Jeandupeux, L. B{\"u}rgi, A. Hirstein, H. Brune, and K. Kern, \prb 59 (1999) 15926.

\bibitem{Ellipse} O. Agam and A. Schiller, \prl 86 (2001) 484.

\bibitem{Ellipse2} G. A. Fiete and E. J. Heller, Rev. Mod. Phys. 75 (2003) 933,
and references therein.

\end{thebibliography}
\end{document}